\documentclass{ws-ijmpa}
\usepackage[super,compress]{cite}
\usepackage{graphicx}
\begin{document}
\markboth{A. Purzsa}{Coulomb interacting Bose-Einstein correlations in Fourier space}

%
\catchline{}{}{}{}{}
%

\title{Coulomb interacting Bose-Einstein correlations in Fourier space}

\author{Aletta Purzsa}
\address{Institute of Physics, E\"otv\"os Lor\'and University, P\'azm\'any P\'eter s\'et\'any 1/A\\
Budapest, H-1117, Hungary\\
purzsaaletta@student.elte.hu}

\maketitle

\begin{history}
\end{history}

\begin{abstract}
In high-energy heavy-ion physics experiments, a state of matter is created that existed in the early Universe: the quark-gluon plasma. This strongly interacting matter exists in today's experiments only within a range of a few femtometers and for a duration of a few femtometers per speed of light, making its resolution with optical tools impossible. However, there is a method that allows for a closer look into the structure of the quark-gluon plasma: femtoscopy. Initially used in astronomy, femtoscopy is based on the quantum mechanical indistinguishability of identical particles, which causes them to arrive at detectors in a correlated manner. The measurable correlation is related to the spacetime structure of the particle-emitting source, which in heavy-ion physics is the quark-gluon plasma created in collisions. For free particles, a relatively simple relationship exists between the source and the correlation (essentially a Fourier transform). However, this relationship becomes complex when accurately accounting for the repulsive Coulomb interaction between final-state electrically charged particles. Our paper presents a new method that is more precise than previously used ones, yet less computationally demanding, especially for the exotic source function shapes. Mathematically, the method is interesting because it exactly handles many emerging integrals and limits. Practically, it is ready for use in experimental analyses.

\end{abstract}



\section{Introduction}
The primary objective of contemporary high-energy nuclear collision experiments is to investigate the strong interaction, with a specific focus on studying a state of matter known as quark-gluon plasma (QGP). Experiments observe the particles produced in heavy ion collisions in order to
draw conclusions about the properties of this matter. The strong interaction is responsible for the attraction between protons and neutrons within the atomic nucleus and was later discovered to also be responsible for the cohesion of the more elementary constituents of protons and neutrons, known as \textit{quarks}.

Quark-gluon plasma is produced during the collisions of atomic nuclei accelerated to relativistic speeds, resulting in the dissociation of protons and neutrons. In this state of matter, quarks and gluons are liberated from nucleons. Under experimental conditions, QGP is produced in particle accelerators, such as the CERN Large Hadron Collider (LHC) and the BNL Relativistic Heavy Ion Collider (RHIC). After its formation, the quark-gluon plasma rapidly expands and cools, and by the time it reaches the detector, it primarily consists of protons, kaons, pions, and other well-known particles.

From the collision data, many interesting observable quantities can be examined, revealing insights into the events of the collision process. An important class of observables is that of momentum correlations of identical particles. In the case of bosonic particles (e.g., pions or kaons), these are called Bose-Einstein-correlations because they arise as a result of the quantum statistical properties of the particles. The fundamental assertion is that the measured correlation functions are related to the source function (i.e., the probability distribution of particle emission), thereby allowing us to infer the spatial-temporal structure of the source at the femtometer scale from the correlation functions. This presents a unique opportunity because, for instance, optical tools cannot reveal such scales in any way. This area of heavy-ion physics, which examines femtometer-scale processes using correlation functions, has evolved into a distinct discipline known as femtoscopy.

In experiments studying heavy-ion collisions, one of the simplest observable quantities is the single-particle momentum distribution, $N_1(\mathbf{p})$, which indicates how many particles are produced in a collision with momenta in an infinitesimal momentum-space volume around $\mathbf{p}$. This distribution can be the summed distribution for the particle type, which is experimentally easier to measure because the detector only needs to measure the particle's momentum, without needing to differentiate the specific type of particle detected. A similar type of quantity is the two-particle distribution  $N_2(\mathbf{p}_1, \mathbf{p}_2)$ , which indicates how many pairs of particles of the same type are produced, where one particle has momentum $\mathbf{p}_1$, and the other has momentum $\mathbf{p}_2$ within small volumes ($\mathrm d^3\mathbf p_1$ and $\mathrm d^3\mathbf p_2$, respectively). In theory, this quantity is measurable, but in practice, we often use the \textit{correlation function} instead. With knowledge of these two quantities, we can introduce the two-particle momentum correlation function (Bose-Einstein correlation function), defined as
\begin{equation}
\label{eq:1}
    C_2 (\mathbf{p}_1, \mathbf{p}_2) = \frac{N_2(\mathbf{p}_1, \mathbf{p}_2)}{N_1(\mathbf{p}_1) N_1(\mathbf{p}_2)} \; .
\end{equation}
In other words, the correlation function is the ratio of the two-particle distribution function to the corresponding single-particle distribution functions. In more intuitive terms, this means how much more likely it is for a particle pair to be created with momenta 
$\mathbf{p}_2$ and 
$\mathbf{p}_2$, compared to if they were created independently with the same momenta. The value of this function is one when there is no correlation between the particle pairs.

The difference of the introduced correlation function 
$C_2$ from one, which characterizes the correlation between observed particles, can be caused by various effects. For instance, secondary particles generated in decays will have correlated momenta. However, in heavy-ion physics, quantum statistical correlations are particularly significant for identical particles.

This type of correlations, initially referred to as ,,intensity correlations," were first observed in astronomy. The credit for this discovery goes to radio astronomers Robert Hanbury Brown and Richard Q. Twiss, who were able to determine the diameter of a distant star using the intensity correlations of incoming photons~\cite{HanburyBrown:1956bqd}. Their results sparked scientific debate, but it was later established that photons arriving from different sources can also arrive at the detector in a correlated manner. This phenomenon is named the HBT effect in their honor. Shortly thereafter, G. Goldhaber, S. Goldhaber, W. Y. Lee, and A. Pais observed similar correlations in elementary particle reactions, specifically on pions produced in proton-antiproton collisions~\cite{Goldhaber:1960sf}. The fundamental explanation was based on the premise that identical particles are indistinguishable in quantum mechanics. Therefore, their wave functions are symmetric for bosons and antisymmetric for fermions with respect to the exchange of the two particles.

\section{Bose-Einstein correlation functions}

Since in quantum mechanics the probability of finding particles is proportional to the square of the absolute value of their wave function ($\psi_{\mathbf{p}} (\mathbf{r})$), where 
$\mathbf{r}$ represents the position and 
$\mathbf{p}$ denotes the momentum of the particle. The single-particle momentum distribution can be obtained if, besides the wave function, we also know the initial distribution function of the quark-gluon plasma produced after the collision, which is called the source function ($S(\mathbf{r}, \mathbf{p})$). Given this information, the single-particle momentum distribution can be calculated as follows:
\begin{equation} \label{eq:eq2}
    N_1(\mathbf{p}) = \int S(\mathbf{r}, \mathbf{p}) \: | \psi_{\mathbf{p}} (\mathbf{r}) | ^2 \: \mathrm{d}^3 \mathbf{r} \; .
\end{equation}
The source function is not an experimentally measurable quantity; we can only infer its shape based on the correlation functions. The simplest assumption for the source function is a Gaussian distribution, and over the past decades, many aspects of Gaussian correlation measurements have been explored. As experimental resolution and data improved over the years, there was a need for a more accurate description of the correlation function's shape~\cite{PHENIX:2006nml}. One approach is using spherical harmonics expansion, while another involves using a source function beyond the Gaussian approximation. Lévy-stable distributions were employed for this purpose, leading to a statistically acceptable description of the data~\cite{Csorgo:2003uv}.

Similarly, the two-particle distribution $N_2(\mathbf{p_1}, \mathbf{p_2})$ can also be obtained using the source functions. Here, $\mathbf{r_1}$ and $\mathbf{r_2}$ represent the positiones, while $\mathbf{p_1}$ and $\mathbf{p_2}$ represent the momenta of particles 1 and 2 respectively. In this case, however, the absolute square of the two-particle wave function needs to be integrated with the source functions corresponding to these positions and momenta as follows:
\begin{equation} \label{eq:eq3}
    N_2(\mathbf{p_1}, \mathbf{p_2}) = \int S(\mathbf{r_1}, \mathbf{p_1}) S(\mathbf{r_2}, \mathbf{p_2}) \: | \psi^{(2)}_{\mathbf{p_1}, \mathbf{p_2} } (\mathbf{r_1,r_2}) | ^2 \: \mathrm{d}^3 \mathbf{r_1} \mathrm{d}^3 \mathbf{r_2} \; .
\end{equation}
This integral combines the source functions $S(\mathbf{r_1}, \mathbf{p_1})$ and $S(\mathbf{r_2}, \mathbf{p_2})$ with the squared magnitude of the two-particle wave function $\psi^{(2)}_{\mathbf{p_1}, \mathbf{p_2} } (\mathbf{r_1,r_2})$.
Due to their quantum mechanical indistinguishability, the wave function for bosons must be symmetrized with respect to the exchange of the two particles. Thus, when taking the absolute square of the wave function, the components of the position and momentum vectors become mixed. Without this symmetrization, the wave function could simply be written as the product of single-particle wave functions, and the value of the correlation function would be a constant one. It is important to note that other aforementioned phenomena can still cause correlations. However, without the quantum statistical properties, studying Bose-Einstein correlations would be meaningless.

\subsection{Case of free particles}
If the two particles are considered free, meaning they arrive at the detector without interaction, their single-particle wave functions can be given in the form of plane waves, whose absolute square is exactly 1. In this case, the symmetrized two-particle wave function can be written as follows, introducing the wave number vectors commonly used in quantum mechanics, $\mathbf{k_1} = \mathbf{p_1} / \hbar$ and $\mathbf{k_2} = \mathbf{p_2} / \hbar$:
\begin{equation} \label{eq:4}
    \psi^{(2)}_{\mathbf{p_1}, \mathbf{p_2} } = \frac{1}{\sqrt{2}} \left(  e^{i \mathbf{k_1} \mathbf{r_1}} e^{i \mathbf{k_2} \mathbf{r_2}} + e^{i \mathbf{k_1} \mathbf{r_2}} e^{i \mathbf{k_2} \mathbf{r_1}} \right) \; .
\end{equation}
To facilitate further calculations, it is also useful to introduce the relative momentum and position vectors, as well as the total momentum and center-of-mass coordinates, along with their magnitudes denoted by the appropriate italic letters:
\begin{align} \label{eq:jelolesek}
\mathbf{k} = \frac{\mathbf{k_1} {-} \mathbf{k_2}}{2} \; , \qquad
\mathbf{r} = \mathbf{r_1} {-} \mathbf{r_2} \; , \qquad
\mathbf{K} = \mathbf{k_1} {+} \mathbf{k_2} \; , \qquad
\mathbf{R} = \frac{\mathbf{r_1} {+} \mathbf{r_2}}{2}    \; . \qquad
\end{align}
In addition, the reduced mass ($m$) and the total mass ($M$) will also be introduced later, in the following sense:
\begin{align}
    m = \frac{m_1 m_2}{m_1 {+} m_2} \;  \qquad \textnormal{és} \qquad
    M = m_1 {+} m_2 \; ,
\end{align}
where $m_1$ and $m_2$ are the masses of the particles. Using these notations, it can be shown after a brief calculation that the absolute square of the two-particle wave function in the non-interacting case can be written as follows:
\begin{equation} \label{eq:5}
   | \psi^{(2)}_{\mathbf{p}_1, \mathbf{p}_2 } (\mathbf{r}) | ^2 = 1 {+} \cos(2 \mathbf{k} \mathbf{r}) \; .
\end{equation}
With this knowledge, equation (\ref{eq:eq3}) can be further transformed by substituting the probability density. According to equation (\ref{eq:eq2}), the first term will include the single-particle momentum distribution associated with the particles. Then, the $\cos(2\mathbf{kr})$ function can be expressed as a sum of exponential functions, leading to the following transformations (c.c. denotes the complex conjugate of the term in square brackets):
\begin{align} \label{eq:atalakitas}
      N_2(\mathbf{p}_1, \mathbf{p_2})  = \int \mathrm{d}^3 \mathbf{r_1} \mathrm{d}^3 \mathbf{r_2} S(\mathbf{r_1}, \mathbf{p_1}) S(\mathbf{r_2}, \mathbf{p_2}) + \int \mathrm{d}^3 \mathbf{r_1} \mathrm{d}^3 \mathbf{r_2} S(\mathbf{r_1}, \mathbf{p_1}) S(\mathbf{r_2}, \mathbf{p_2})  \cos(2 \mathbf{kr}) =  \nonumber \\ 
      =  N_1(\mathbf{p_1})N_1(\mathbf{p_2}) + \frac{1}{2} \left[ \int \mathrm{d}^3 \mathbf{r_1} e^{i 2 \mathbf{kr_1}} S(\mathbf{r_1}, \mathbf{p_1}) \int \mathrm{d}^3 \mathbf{r_2} e^{-i 2 \mathbf{kr_2}} S(\mathbf{r_2}, \mathbf{p_2}) + \mathrm{c}. \mathrm{c}. \right] \; . 
\end{align} 
For particle pairs produced in a correlated manner, we can make the approximation based on experimental data that in the source function of the particles, $\mathbf{p_1} \approx \mathbf{p_2} \approx \mathbf{K} / 2$. However, we do not make this approximation in the exponential factor. Performing this transformation introduces the Fourier transform of the $S$ function, in the following sense:
\begin{equation}
    \tilde{S}(2 \mathbf{k} , \mathbf{K}/2) := \int \mathrm{d}^3 \mathbf{r} e^{i 2\mathbf{kr}} S(\mathbf{r}, \mathbf{K}/2) \; .
\end{equation}
Thus, the two-particle momentum distribution and the correlation function can be expressed in the following form:
\begin{equation}
    N_2(\mathbf{p_1}, \mathbf{p_2}) = N_1(\mathbf{p_1})N_1(\mathbf{p_2}) + | \tilde{S}(2 \mathbf{k}, \mathbf{K} / 2) | ^2 \; ,
\end{equation}
\begin{equation}
    C_2(\mathbf{p_1}, \mathbf{p_2}) \equiv C(\mathbf{k}, \mathbf{K}) \cong 1 + \frac{| \tilde{S}(2 \mathbf{k}, \mathbf{K}/2) |^2}{| \tilde{S}( \mathbf{0}, \mathbf{K}/2 )|^2} \; .
\end{equation}
This means, that from now on, the correlation function will be expressed as a function of the relative momentum ($\mathbf{k}$) and the total momentum ($K$). This is identically equal to the two-particle correlation function $C_2(\mathbf{p_1}, \mathbf{p_2})$, but redefined in terms of these new momentum variables for simplicity and clarity in further analysis. The denominator of the fraction is obtained from the product 
$N_1(\mathbf{p_1})N_1(\mathbf{p_2})$, using the same approximation $\mathbf{p_1} \approx \mathbf{p_2} \approx \mathbf{K} / 2$.

The above statements are only valid if, in addition to both particles being bosons, there is no interaction between them. In the presence of interactions, the wave function will take on a more complex form, making the transformation performed in equation (\ref{eq:atalakitas}) no longer applicable.

\subsection{The effect of the Coulomb Interaction}
In practice, a significant number of particles produced in heavy-ion physics experiments are electrically charged. In such cases, the assumption that particles arrive at the detector without interaction does not hold. Therefore, we cannot use the formula derived for free particles; instead, the Coulomb interaction between the particle pairs must also be taken into account. Since the particle pairs consist of identical particles, their mass ($m_1 = m_2$) and atomic number ($z_1 = z_2 = z$) are the same. The Sommerfeld parameter, denoted by $\eta$, which also depends on the particle's momentum, is introduced:
\begin{equation}
    \eta := z_1 z_2 \frac{q_e^2}{4 \pi \epsilon_0} \frac{1}{\hbar c} \frac{mc^2}{pc} 
\end{equation}
The single- and two-particle momentum distributions can still be expressed as in equations (\ref{eq:eq2}) and (\ref{eq:eq3}), respectively. However, due to the presence of the interaction, the wave function corresponding to the Coulomb interaction must be used instead of the plane wave. For the two-particle momentum distribution, the two-particle wave function related to the relative motion should be used, which can be written in the following form~\cite{Landau}:
\begin{equation} \label{eq:coulomb_wf}
    \psi_{\mathbf{k}} (\mathbf{r}) = \frac{1}{\sqrt{2}} \mathcal{N} e^{i \mathbf{KR}} e^{- ikr} \Bigl[ M(1{-}i \eta, 1, i(kr {+} \mathbf{kr})) +M(1{-}i \eta, 1, i(kr {-} \mathbf{kr})) \Bigr] \; .
\end{equation}
where $\mathcal{N}$ is a normalization factor, whose absolute square is known as the \textit{Gamow factor}:
\begin{equation}
    \mathcal{N} = e^{- \pi \eta / 2} \Gamma(1{-}i \eta) \qquad \Rightarrow  \qquad  |\mathcal{N}|^2 = \frac{2 \pi \eta}{e^{2 \pi \eta} {-} 1} \; .
 \end{equation}
Since the value of $\eta$ depends on the momentum, the Gamow factor must be calculated separately for each momentum value. In the definition, $\Gamma(z)$ denotes the gamma function, and $M(a, b, z)$ is the confluent hypergeometric function, which can be expressed using its infinite series for any $a, b, z \in \mathbb{C}$
\begin{equation} \label{eq:hypergeometric}
     M(a, b, z) = \sum_{n = 0}^{\infty} \frac{\Gamma(a {+} n)}{\Gamma(b {+} n)} \frac{\Gamma(b)}{\Gamma(a)} \frac{z^n}{n!} = 1 + \frac{a}{b} \frac{z}{1!} \frac{a(a{+}1)}{b(b{+}1)} \frac{z^2}{2!} + \frac{a(a{+}1)(a{+}2)}{b(b{+}1)(b{+}2)} \frac{z^3}{3!} + \cdots \; .
\end{equation}
One of the simplest methods for accounting for the Coulomb interaction is known as the Gamow method. The essence of this approach is to multiply the correlation function, which is expressed using plane waves, by the (momentum-dependent) Gamow factor. This provides a good approximation for small source sizes. For more extensive source functions, however, a more accurate result is obtained by calculating the correlation function based on the Koonin-Pratt formula~\cite{Koonin:1977fh}.
With this understanding, the aim is to calculate the correlation function, which is carried out in the following manner:
\begin{equation} \label{eq:corr_function}
     C(\mathbf{k} ) = \int \mathrm{d}^3 \mathbf{r} \: D(\mathbf{r}) | \psi_{\mathbf{k}} (\mathbf{r}) |^2 \; .
 \end{equation}
 The function $D(\mathbf{r})$, which appears here and is known as the pair distribution function in coordinate space, can be written as follows:
\begin{equation}
    D(\mathbf{r}) = \int \mathrm{d}^3 \mathbf{R} \; S \left(\mathbf{R} {+} \frac{\mathbf{r}}{2}, \frac{ \mathbf{K} }{2}  \right) \; S \left(\mathbf{R} {-} \frac{\mathbf{r}}{2}, \frac{ \mathbf{K} }{2} \right) \; .
\end{equation}
The Gaussian distribution has been the simplest assumption for the source function, and extensive research over the past decades has explored many aspects of Gaussian correlation measurements~\cite{Csanad:2008af}. However, with the advancement in experimental resolution and the increasing volume of data, a more accurate description of the correlation function's shape became necessary~\cite{PHENIX:2006nml}. Lévy-stable distributions have been used for this purpose, providing a statistically successive way to describe the measurement data~\cite{Csanad:2024jpy,Csanad:2024hva} . As these more sophisticated source functions, like Lévy distributions, become more widely adopted, the need for precise computational tools to manage the Coulomb interaction has become increasingly important.
The Lévy distribution is defined as follows:
\begin{align}
 \label{eq:levy}
\mathcal{L}(\alpha , R, \mathbf{r}) = \frac{1}{(2 \pi) ^3} \int \mathrm{d}^3 \mathbf{q} \; e^{- \frac{1}{2} |\mathbf{q} R|^{\alpha}} e^{i \mathbf{q} \mathbf{r}} \; ,
\end{align}
Where $\alpha$ is a parameter of the distribution that determines its shape. The value of $\alpha$ is typically constrained to the range $0 < \alpha \leq 2$, with $\alpha = 2$ corresponding to a Gaussian distribution. The parameter $R$ represents the characteristic radius of the distribution, and $\mathbf{r}$ is the spatial variable.

One can show that if the source function can be expressed in the form of a Lévy distribution with radius $R$, then the pair distribution function can also be given in the form of a Lévy distribution with radius $2^{\frac{1}{\alpha}}R$, as follows:
\begin{equation}
    D(\mathbf{r}) = \mathcal{L}( \alpha, 2^{\frac{1}{\alpha}} R, \mathbf{r}) \; ,
\end{equation}

Given this information, the correlation function can be obtained by first numerically calculating the Lévy source function. Then, the correlation function must be determined using equation (\ref{eq:corr_function}) through an additional numerical integration, which is a very time-consuming process due to the slow decay of the Lévy distribution. The objective of this paper was to find a method that could overcome this lengthy and highly computationally demanding process.

\section{Novel method for calculating the Coulomb interaction}
The goal of the calculation is to determine the correlation function $C(\mathbf{k})$ based on equation (\ref{eq:corr_function}). For further transformations, it is crucial that the pair distribution function $D(\mathbf{r})$ is integrable and can be expressed as the Fourier transform of another integrable function $f(\mathbf{q})$.  In practice, since this function follows a Lévy distribution, this condition will be met, that is:
\begin{align} \label{eq:pareloszlas}
      D(\mathbf{r}) = \int \frac{\mathrm{d}^3 \mathbf{q}}{(2 \pi)^3} f(\mathbf{q}) e^{i \mathbf{qr}} \qquad \Leftrightarrow  \qquad  f(\mathbf{q}) = \int \mathrm{d}^3 \mathbf{r} \, D(\mathbf{r}) e^{-i \mathbf{qr}} \; .
\end{align}
Given this, we can substitute the pair distribution function expressed through its Fourier transform into the right side of equation (\ref{eq:corr_function}):
\begin{align}
    C(\mathbf{k}) = \int \mathrm{d}^3 \mathbf{r} \int \frac{\mathrm{d}^3 \mathbf{q}}{(2 \pi)^3} f(\mathbf{q}) e^{i \mathbf{qr}} |\psi _{\mathbf{k}}(\mathbf{r})|^2  \; .
\end{align}
The goal is to interchange the order of the two integrations, which can be accomplished by introducing a \textit{regularization factor} $e^{- \lambda r}$ with a parameter $\lambda \in \mathbb{R}^+$. In this way, the necessary transformations can be performed, and ultimately, taking the limit $\lambda \to 0$. Specifically, this means that:
\begin{align}
     \int \mathrm{d}^3 \mathbf{r} \int \frac{\mathrm{d}^3 \mathbf{q}}{(2 \pi)^3} f(\mathbf{q}) e^{i \mathbf{qr}} |\psi _{\mathbf{k}}(\mathbf{r})|^2 \qquad \Rightarrow  \qquad  \int \frac{\mathrm{d}^3 \mathbf{q}}{(2 \pi)^3} f(\mathbf{q}) \int \mathrm{d}^3 \mathbf{r} \: e^{i \mathbf{qr}} |\psi_{\mathbf{k}} (\mathbf{r})|^2 \; .
\end{align}
In other words, the goal is to first perform the Fourier transformation of $| \psi_{\mathbf{k}}(\mathbf{r})|^2$ and then apply this result to $f(\mathbf{q})$. To achieve this, a regularizing factor is introduced. The condition is satisfied that $\psi_{\mathbf{k}}$ is a bounded function and $D(\mathbf{r})|\psi_{\mathbf{k}}(\mathbf{r})|^2$ is integrable, allowing the following transformations.  In the first step, due to the Lebesgue theorem, the limit can be moved outside the integral. In the second step, applying the Fubini theorem allows for interchanging the order of integration:
\begin{align} \label{eq:fubini}
    C(\mathbf{k}) \stackrel{1.}{=} \lim_{\lambda \to 0} \int_0^{\infty} \mathrm{d}^3 \mathbf{r} \: e^{-\lambda r} |\psi_{\mathbf{k}} (\mathbf{r})|^2  \int \frac{\mathrm{d}^3 \mathbf{q} }{(2 \pi)^3 } f(\mathbf{q}) e^{i \mathbf{q} \mathbf{r}} \stackrel{2.}{=} \nonumber \\
    \stackrel{2.}{=} \lim_{\lambda \to 0} \int \frac{\mathrm{d}^3 \mathbf{q}}{(2 \pi)^3} f(\mathbf{q}) \int \mathrm{d}^3 \mathbf{r} \: e^{i \mathbf{q} \mathbf{r}} e^{- \lambda r} |\psi_{\mathbf{k}}(\mathbf{r})|^2 \: .
\end{align}
Obviously, the $\lambda \to 0$ cannot be moved inside the integrals. From this point on, it is assumed that the function $f$ is spherically symmetric, and this case will be distinguished with the index \textit{s}, that is the correlation function will be denoted as $C_s(\mathbf{k})$, and the function $f$ will be denoted as $f_s$. These then depend only on $k$ (the magnitude of $\mathbf{k}$) and $q$ (the magnitude of $\mathbf{q}$) and are denoted as $C_s(k)$ and $f_s(q)$ respectively. In this case, the integral over the solid angle $S^2$ can be performed, leading to the following relationships ($\mathbf{n}$ denotes the direction of the vector $\mathbf{q}$):
\begin{align}
    C_s(\mathbf{k}) = \lim_{\lambda \to 0} \int_{0}^{\infty} \mathrm{d} q \frac{q^2}{8 \pi ^3} f_s(q) \int \mathrm{d}^3 \mathbf{r} e^{- \lambda r} |\psi_{\mathbf{k}} (\mathbf{r})|^2 \int_{S^2} \mathrm{d} \mathbf{n} \: e^{i q\mathbf{nr}} = \\
     = \lim_{\lambda \to 0} \int_{0}^{\infty} \mathrm{d} q  f_s(q) \int \mathrm{d}^3 \mathbf{r} \frac{e^{- \lambda r}}{2 \pi ^2 r} |\psi_{\mathbf{k}}(\mathbf{r})|^2 \sin(qr) \nonumber \; .
\end{align}
The two-particle wave function corresponding to relative motion, as described in equation (\ref{eq:coulomb_wf}), can be substituted. Upon doing this, it follows that:
\begin{align} \label{eq:lim_out}
    C_s(k) = \frac{| \mathcal{N} |^2}{2 \pi ^2} \lim_{\lambda \to 0} \int_0^{\infty} \mathrm{d} q \: q^2 f_s(q) \Bigl[ \mathcal{D}_{1 \lambda s} (q) + \mathcal{D}_{2 \lambda s} (q) \Bigr] \: ,
    \end{align}
    where
    \begin{align}
    \mathcal{D}_{1 \lambda s} (q) = \hspace{-2pt}\int \hspace{-3pt}\mathrm{d} ^3 \mathbf{r} \frac{\sin(qr)}{qr} e^{- \lambda r} M \bigl(1{+}i \eta, 1, -i(kr {+} \mathbf{kr}) \bigr) \: M \bigl(1{-}i \eta, 1, -i(kr {+} \mathbf{kr}) \bigr) \; \quad 
    \end{align}
    and
    \begin{align}
    \mathcal{D}_{2 \lambda s} (q) = \int \mathrm{d} ^3 \mathbf{r} \frac{\sin(qr)}{qr} e^{- \lambda r} M \bigl(1{+}i \eta, 1, -i(kr {-} \mathbf{kr}) \bigr) \: M \bigl(1{-}i \eta, 1, -i(kr {+} \mathbf{kr}) \bigr) \; .
\end{align}
$\mathcal{D}_{1 \lambda s} (q)$ and $\mathcal{D}_{2 \lambda s} (q)$ 
can be calculated using the method described in Ref.[~\citen{Nordsieck:1954zz}].
The obtained $\mathcal{D}_{1 \lambda s} (q)$ and $\mathcal{D}_{2 \lambda s} (q)$ functions are not integrable in the domain $ q \in [0, \infty [ $ as $\lambda \to 0$, and thus, the Lebesgue theorem cannot be applied. The trick is that by subtracting and then adding back $f_s(0)$ and $f_s(2k)$ in the result, the outcome will become integrable.
The subsequent calculations are not covered within the scope of this paper; they can be found in Ref.[~\citen{Nagy:2023zbg}].

The final result after the calculations:
\begin{align} \label{eq:korr}
    C_s(k) = |\mathcal{N}|^2 \biggl[ f_s(0) + f_s(2k) + \frac{\eta}{\pi} (\mathcal{A}_{1s} + \mathcal{A}_{2s}) \biggr] \; ,
\end{align}
where
\begin{align} \label{eq:a1s}
    \mathcal{A}_{1s} = - \frac{2}{\eta} \int_{0}^{\infty}\mathrm{d}q \frac{f_{s}(q) {-} f_s(0)}{q} \, \textnormal{Im} \Bigl[ \left(1 {+} 2k/q \right)^{2 i \eta} \mathcal{F}_+ (4k^2 / q^2 -i0) \Bigr] \; ,
\end{align}
and
\begin{align} \label{eq:a2s}
    \mathcal{A}_{2s} = - \frac{2}{\eta} \int_0^{\infty} \mathrm{d}q \frac{f_s(q){-}f_s(2k)}{q{-}2k} \frac{q}{q{+}2k} \textnormal{Im} \frac{(q{+}2k)^{i \eta}}{(q{-}2k{+}i0)^{i \eta}} \; 
\end{align}
and
\begin{align}
\mathcal{F}_+ (x) :=  \; _2F_1 (i \eta, 1{+}i \eta, 1, x) \; ,
\end{align}
where $_2F_1$ is the Gaussian hypergeometric function \cite{NIST:DLMF}.

\begin{figure}[h!]
    \centerline{
    \includegraphics[width=0.5\textwidth]{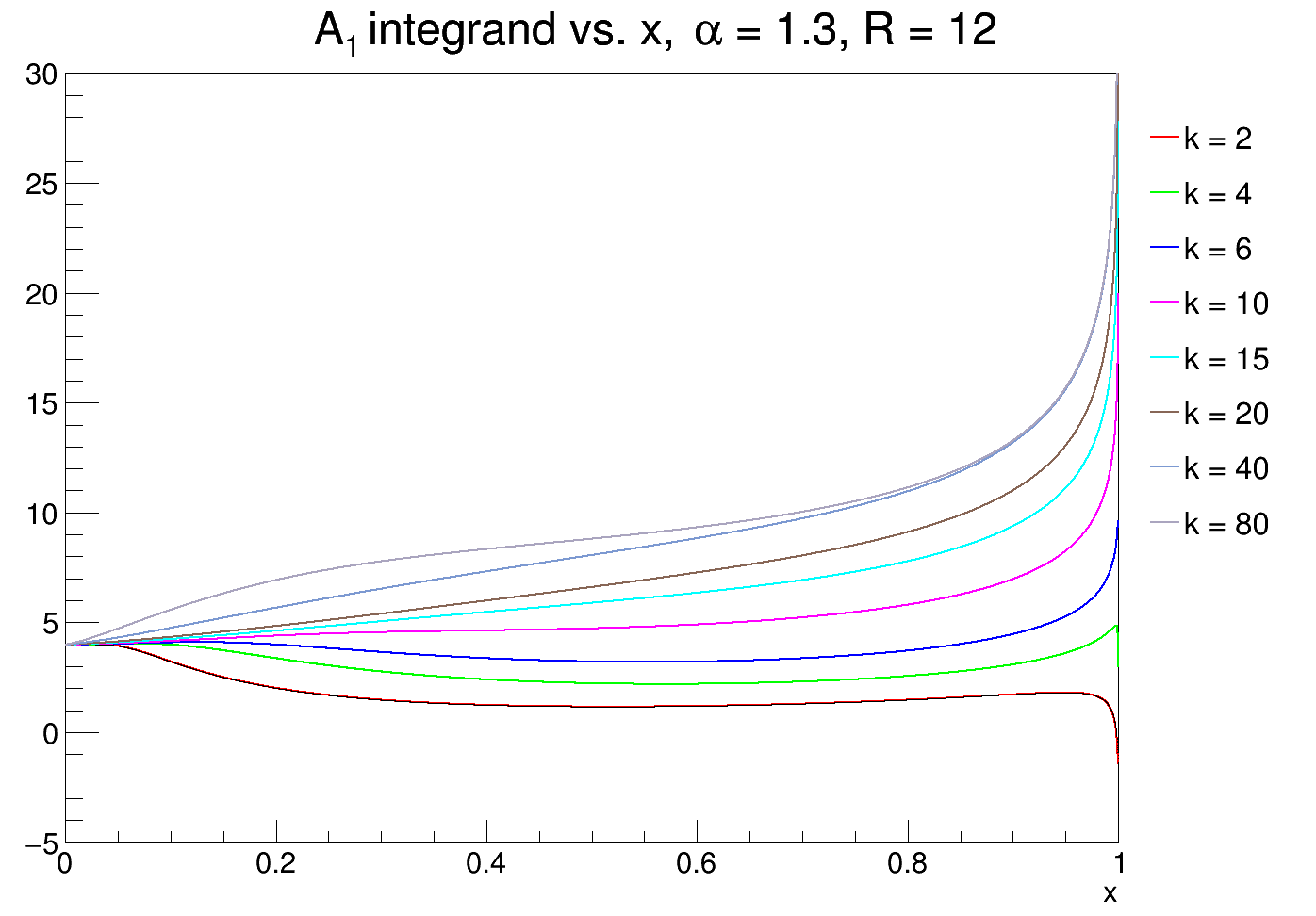}\includegraphics[width=0.5\textwidth]{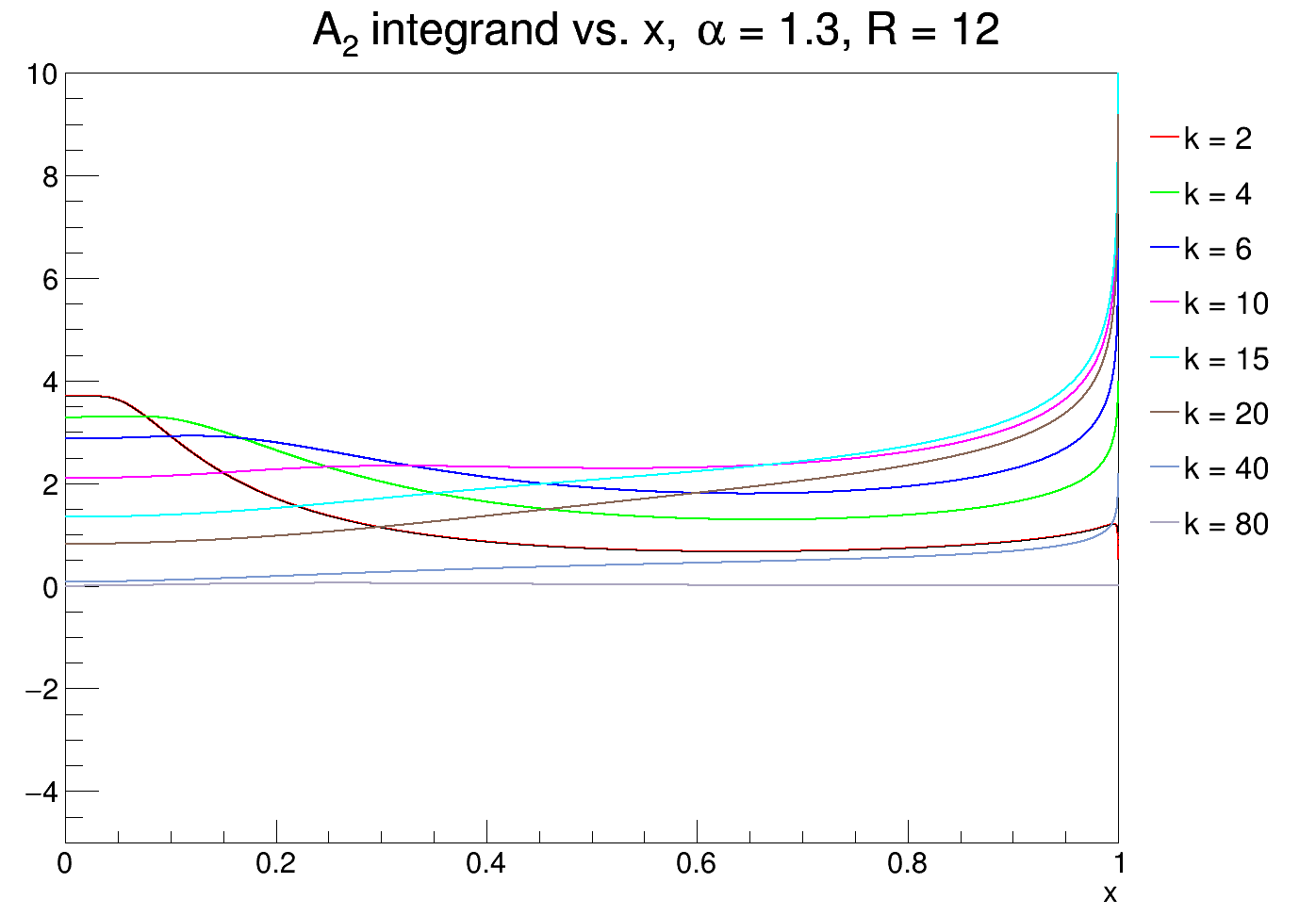}}
    \caption{$\mathcal{A}_{1s}$ and $\mathcal{A}_{2s}$  integrands plotted for $\alpha = 1.3$ and $R = 12$. The graphs show that the functions are smooth, making them easily integrable. The only critical point is at $x = 1$ where the function becomes logarithmically oscillatory, but it remains bounded, so this does not affect integrability.}
    \label{fig:integrands}
\end{figure}

\begin{figure}[h!]
    \centerline{
    \includegraphics[width=0.5\textwidth]{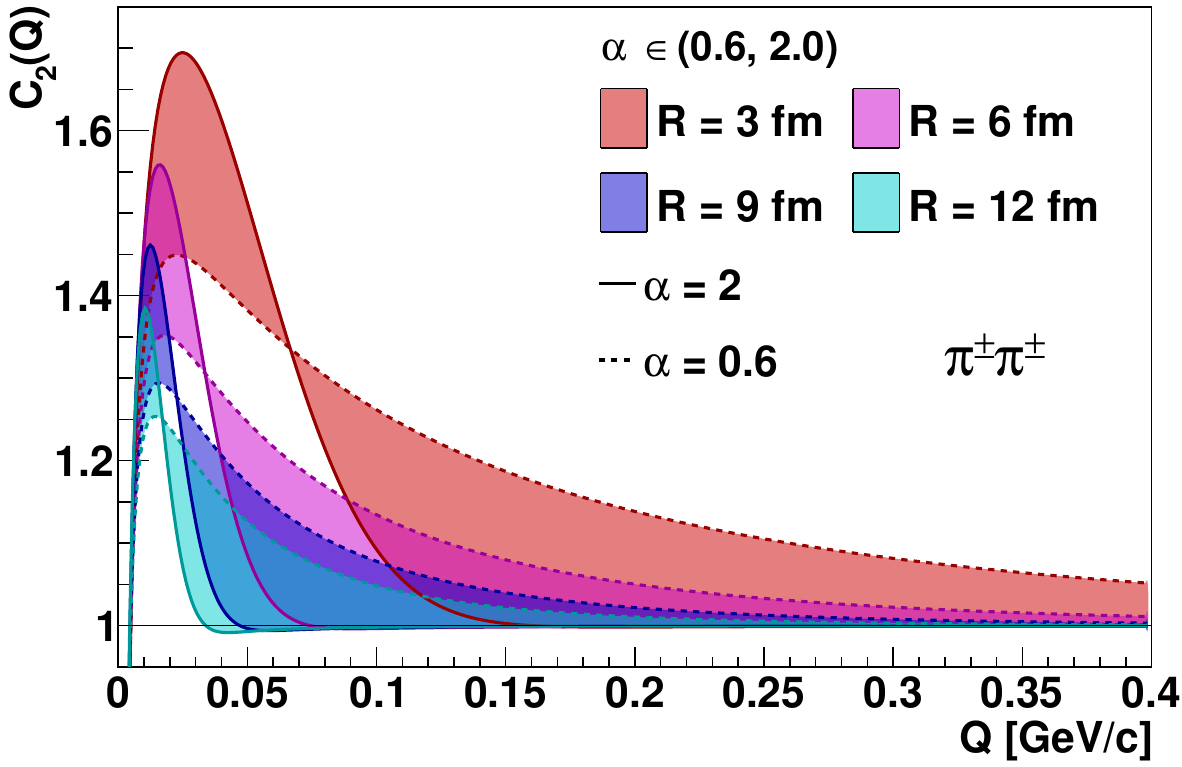}\includegraphics[width=0.5\textwidth]{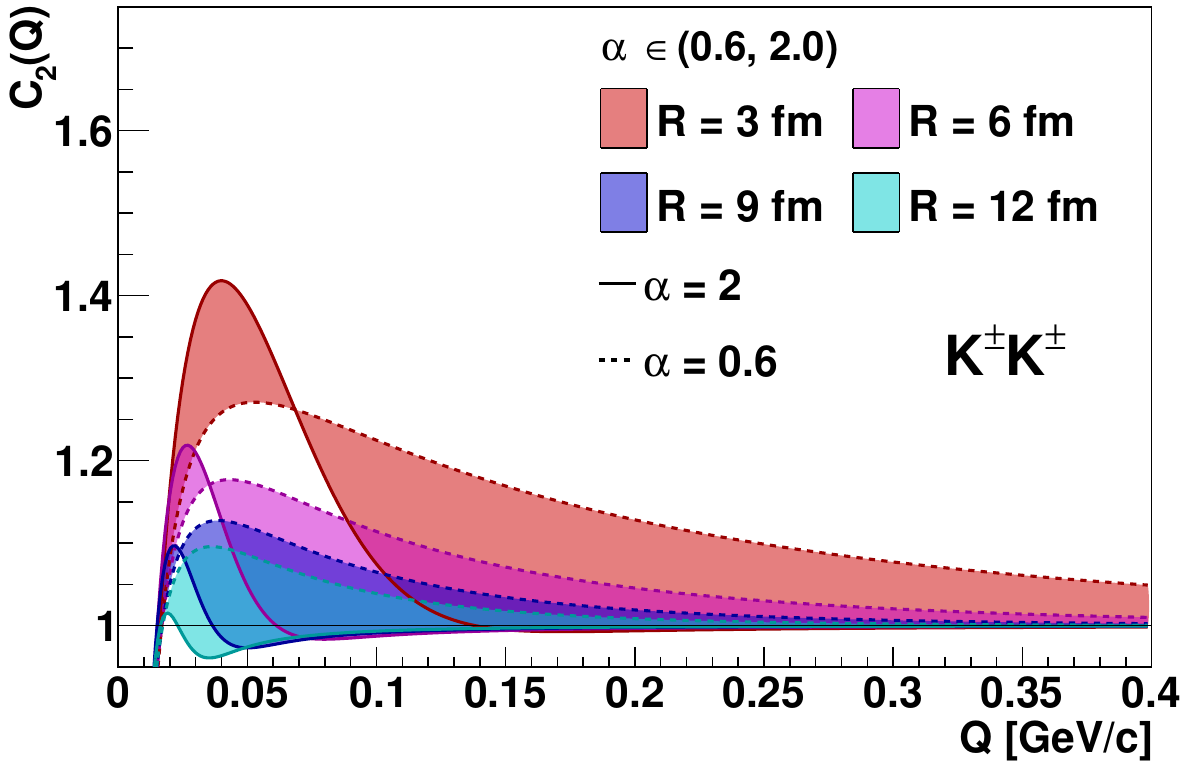}}
    \caption{Example correlation functions for pions (left) and kaons (right), plotted for four different $R$ and two $\alpha$ values. $Q = 2 k$, and the $K$ argument is dropped for simplicity. At a given $R$ value, the shape of the correlation function with increasing $\alpha$ values from $\alpha\,{=}\,0.6$ to $\alpha\,{=}\,2$ goes smoothly through the shaded region.}
    \label{fig:compare}
\end{figure}

\begin{figure}
    \centerline{
    \includegraphics[width=0.5\textwidth]{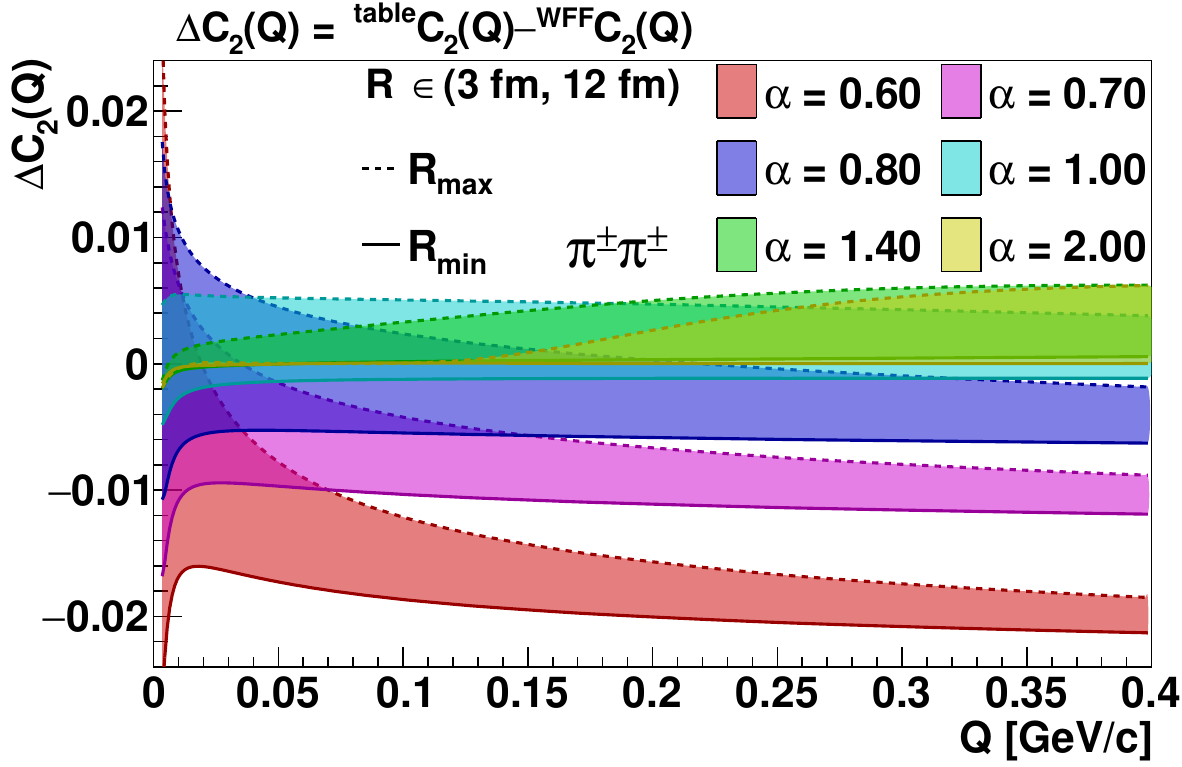}\includegraphics[width=0.5\textwidth]{ijmpa-2e/compare_kaoncorr.pdf}}
\caption{Difference between the correlation function calculated with a numerical integral method described in Ref.~\citen{Kincses:2019rug} ($^{\textnormal{table}}C_2(Q)$) and the correlation function calculated with the wave-function Fourier method described in the current paper ($^{\textnormal{WFF}}C_2(Q)$). $\Delta C_2(Q)$ is plotted for 6 different $\alpha$ values and two $R$ values, for pions (left) and kaons (right) separately. At a given $\alpha$ value, $\Delta C_2(Q)$ goes smoothly through the shaded region when increasing $R$ values from $R = 3\textnormal{ fm}$ to $R = 12\textnormal{ fm}$.}
    \label{fig:subtr}
\end{figure}

Figures \ref{fig:integrands}, \ref{fig:compare}, and \ref{fig:subtr} collectively illustrate the behavior and consistency of the integrands and correlation functions under varying parameters. 

\section{Summary}
In summary, the goal was to develop a more efficient method than the previously used Coulomb correction procedure to support measurements assuming a Lévy distribution for the source function. The key element of the method is first performing the Fourier transform of the interacting two-particle wave function and then applying this resulting distribution to the Fourier transform of the source function. Our method (described in more detail in Ref ~\citen{Nagy:2023zbg}) allows for much more efficient calculation of correlation functions compared to the previously used approach. Previously, after calculating the source function, a numerical integration had to be performed, which was computationally intensive and took days for the program to run. The results then had to be stored in a memory-intensive table, which needed to be loaded for later use. With the newly described method, however, correlation functions can be calculated in just a few minutes, even using a desktop computer.

The presented method is only applicable only to the spherically symmetric case. However, a more general procedure exists, and its numerical implementation is a objective for the future.

A software package is also published that is easily applicable and ready to use in experimental analyses \cite{software}.

\section{Acknowledgments}
This work was supported by the Hungarian NKFIH grants TKP2021-NKTA-64 and K-138136.


\end{document}